# Prolonged photo-carriers generated in a massive-and-anisotropic Dirac material

Munisa Nurmamat[1,*], Yukiaki Ishida[2], Ryohei Yori[1], Kazuki Sumida[1], Siyuan Zhu[1],

Masashi Nakatake[3], Yoshifumi Ueda[4], Masaki Taniguchi[4], Shik Shin[2], Yuichi Akahama[5],

and Akio Kimura[1,¶]

[1]Department of Physical Sciences, Graduate School of Science, Hiroshima University,
1-3-1 Kagamiyama, Higashi-Hiroshima 739-8526, Japan
[2]Institute for Solid State Physics, the University of Tokyo,
5-1-5 Kashiwa-no-ha, Kashiwa, Chiba 277-8581, Japan
[3]Aichi Synchrotron Radiation Center, Aichi Science & Technology Foundation,
250-3 Minamiyamaguchi-cho, Seto 489-0965, Japan
[4]Hiroshima Synchrotron Radiation Center, Hiroshima University,
2-313 Kagamiyama, Higashi-Hiroshima 739-0046, Japan
[5]Graduate School of Material Science, University of Hyogo,
3-2-1 Kouto, Kamigori-cho, Ako-gun, Hyogo, Japan
*e-mail: munisa627@hiroshima-u.ac.jp    ¶e-mail: akiok@hiroshima-u.ac.jp

**Abstract**

Transient electron-hole pairs generated in semiconductors can exhibit unconventional excitonic condensation. Anisotropy in the carrier mass is considered as the key to elongate the life time of the pairs, and hence to stabilize the condensation. Here we employ time- and angle-resolved photoemission spectroscopy to explore the dynamics of photo-generated carriers in black phosphorus. The electronic structure above the Fermi level has been successfully observed, and a massive-and-anisotropic Dirac-type dispersions are confirmed; more importantly, we directly observe that the photo-carriers generated across the direct band gap have the life time exceeding 400 ps. Our finding confirms that black phosphorus is a suitable platform for excitonic condensations, and also open an avenue for future applications in broadband mid-infrared BP-based optoelectronic devices.

## Introduction

Formation of electron-hole (e-h) pairs called excitons has renewed interest of research since the BCS (Bardeen-Cooper-Schrieffer) like e-h Cooper pairs condensation is expected in semiconductors [1-2]. By optical pumping, photoexcited carriers facing across the band gap form excitons due to the Coulomb interaction between electrons and holes with low carrier density [3]. These excitons give rise to an unconventional excitonic state in direct band gap semiconductors and hence arouse much interest of pump-probe studies currently [4]. The exciton condensation has been predicted theoretically [1-3, 5] and afterwards realized experimentally [6-9], but the decisive evidence of excitonic condensation in optically pumped semiconductor is elusive. Recently, it is theoretically suggested that e-h BCS order can be enhanced when the effective masses of electrons and holes have a strong anisotropy [10-11]. Besides, the photo-induced superconductivity has also been predicted in a laser-driven two-band semiconductor [12]. These studies encourage us to challenge experimental observation of excitonic condensation in a promising two-dimensional (2D) semiconductor possessing both direct band gap and high effective mass anisotropy.

Here, we focus on orthorhombic black phosphorus (BP), which is the most stable allotrope of phosphorus with a direct band gap of ∼ 0.3 eV. It forms a folded honeycomb sheet running along the *a*-axis as shown in Fig. 1(a). One of the unique features distinguishing BP from other 2D materials is its anisotropic transport properties. The hole mobility has been reported to reach about 3000 $cm^2V^{-1}s^{-1}$ at 200 K for bulk crystal [13] and 1000 $cm^2V^{-1}s^{-1}$ for 15nm-thick film along *c*-axis [14], which are more than 1.8 times larger than that along *a*-axis. Such anisotropic nature of BP allows us to build a high-performance transistor realizing a ballistic transport which are strongly related to the anisotropic effective mass. The effective masses of holes and electrons along three axes have been studied by far-infrared cyclotron resonance [15]. The tunable band gap structure of BP determined by the number of layers from 0.3 eV (bulk) to 2 eV (single monolayer) make BP as a promising nonlinear optical material, particularly with great potentials for infrared and mid-infrared optoelectronics [16-19]. As being the direct band gap and high anisotropic 2D semiconductor, BP can be expected to form excitons due to the nonequilibrium electron and hole population after optical pumping and hence to condensate the excitonic e-h state. Very recently, several research groups have studied the carrier dynamics and anisotropic dynamical response in bulk and few layer BP by using pump-probe transmission and reflection measurements [20-24]. Furthermore, angle resolved photoemission spectroscopy (ARPES) have so far been applied to get an information of hole bands (valence band) of bulk BP [25-27] as well as its band gap variation by alkali metal doping [28-29]. However, the electrons (holes) excited in the conduction (valence) band and their carrier dynamics have not yet been clarified.

Time- and angle-resolved photoemission spectroscopy (TARPES), the conventional ARPES implemented by a pump-and-probe method, is a powerful tool to study the electron/hole bands and electron/hole dynamics with energy and momentum resolutions. Recent TARPES studies on three-dimensional topological insulators have demonstrated that the photoexcited electrons are bottlenecked at the Dirac point of the topological surface state leading to the population inversion up to ∼3 ps [30] and exhibiting an electronic recovery time variation from ∼5 ps to ∼400 ps by the closeness of $E_F$ to the Dirac point as well as upon the increase in the bulk insulation [31]. Thus, to understand population of photoexcited electrons (holes) in the conduction (valence) band and its carrier dynamics of BP, which offer the hints on e-h pairs condensation, the TARPES is essential. Here we show the dynamics of photo-generated electrons/holes as well as their anisotropic features utilizing the combined tools of ARPES and two-photon photoemission spectroscopy. More interestingly, we have directly observed the population of electrons and holes across the direct band gap that persistently remain with a long relaxation time of more than 400 ps.

**Results and discussions**

**Sample characterization.** The high-quality surface properties of bulk BP have been characterized by scanning tunneling microscopy (STM). Fig. 1(c) shows the representative STM topographic image on freshly cleaved BP surface with additional defects which can be understood as impurities/or vacancies within BP crystals. The zoomed-up image with atomic resolution shows more detailed surface structure of BP along *a*- and *c*-directions as described in Fig. 1(d). The in-plane lattice constant along *c* (lighter effective mass) - and *a* (heavier effective mass)-directions are estimated from our STM image as 4.3 and 3.34 Å, and it is consistent with previous works [32-34].

**Hole band and their effective mass.** In order to investigate the dispersion of valence band maximum at Z point of the bulk Brillouin zone [see Fig. 1(b)], the normal emission spectra are taken with several photon energies (h$\nu$=19-56 eV). The dispersive feature has been clearly observed especially near $E_F$ [see Fig. S1 of Supplementary Material]. Eventually, we find that the valence band is closest to $E_F$ at two photon energies of 19 and 52 eV. Since the photoemission intensity is stronger at 19 eV rather than at 52 eV, we choose 19 eV to take the in-plane ARPES dispersion at the valence band maximum.

Figures 2(a-b) demonstrate the in-plane ARPES dispersion curves along *c*- and *a*-axes. Both results show the downward dispersion near $E_F$, while the upward dispersions are also observed on the higher $E_B$ side. Now we focus on the band showing the downward valence band dispersion. To estimate the effective masses along *c*- and *a*-axes, we have first fitted the results with the polynomial function. After that, we calculated the in-plane hole effective mass *m\** by using the relation $h^2/m^* = d^2E/dk^2$, where $\hbar$ is Planck's constant and $k$ is the wave number

corresponding to the momentum. Finally, we obtain $m_a^*/m_0 = -0.54$ and $m_c^*/m_0 = -0.05$, which means that the effective mass along *a*-axis is about 10 times larger than that along *c*-axis. These values are a little smaller than the cyclotron mass; -0.648 and -0.076, respectively [15].

**Electron band and its effective mass observed by TARPES.** Now we turn to the conduction band, which can be directly accessed by TARPES. Figures 3(a-b) show TARPES images taken along the *c*- and *a*- axes recorded before (*t* = -1.33 ps) and after (*t* = 1.06 ps) pump. Before pumping, there is no photoelectron intensity from the conduction band [see left panels of Figs. 3(a-b)] and only the valence band dispersion obtained here along *c*- and *a*- axes, again signifies its anisotropic nature as discussed above. The unoccupied states are filled after the pumping and form the upward parabolic band dispersion along *c*-axis with the energy minima at ∼ 0.3 eV as shown in right-panel of Fig. 3(a). In a sharp contrast, the energy-band dispersion along *a*-axis observed at the same delay time is much more flattened [see right-panel of Fig. 3(b)]. These obvious shape-different band dispersions along *c*- and *a*- axes signifies a giant anisotropy in the conduction band of BP. The bottom of the conduction band is located farther from $E_F$ than the top of the valence band with the assumed direct energy gap of 0.32 eV showing a p-type semiconductor sign. We have estimated the effective mass with the same procedure as those for the ARPES. As a result, the effective masses along *c*- and *a*-axes are 0.047 and 1, respectively.

**Dynamics of pump generated carriers.** To unveil the dynamics of pump-generated carriers in the unoccupied state, we present the transient ARPES spectra with the typical pump-and-probe delay times. The upper panels of Fig. 4(a) show TARPES images taken along the ΓZLX plane (*c*-axis) recorded after (0 ≤ *t* ≤ 475 ps) pump whereas the lower panels show their difference image to that before pumped. The population intensity of electrons (holes) in the conduction (valence) band that was increased (decreased) as a function of delay times are more clearly reflected in lower panels. The valence band electrons are excited to the unoccupied state immediately after the pump and cascade from the high-energy region to the bottom of conduction band as time delays. From the difference spectra, we have distinctly seen that the massive Dirac-type dispersion formed by the excited electrons and holes across the direct band gap *i.e.* the holes accumulating at the top of the valence band and facing the electrons piling up at the bottom of the conduction band. In the decaying process, the electrons in the unoccupied states lose its energy quickly in the range of *t* = 0 ∼ 3 ps. In contrast, the electrons in the bottom of conduction band exhibit a long relaxation time more than 400 ps. Note that it is much longer than that was observed for graphene up to ∼ 1 ps [35-36].

In order to discuss the energy-dependent dynamics, we provide the time evolution of the intensity integrated in the energy and momentum frames A to D [see Fig. 4(b)]. The intensity is normalized by the peak amplitude of each frame. In a few picoseconds after pumping, the intensity rises steeply and then it decreases exponentially. By comparing these frames, the comparative fast decay is found in frames A and B, these may be considered as the photon-induced interband scattering towards the lower conduction band which is similar to the graphene [37]. On the contrary, the decay process becomes slower in frames C and D, and also the intensity persists long systematically [particularly in Frame D]. For the relatively slow decay in frames C and D, after the electrons piled up to the conduction band bottom, the recombination with the holes in the valence band will occur due to the narrow semiconducting gap.

We further quantify the decay process by investigating the profile of the dissipation. To this end, we derive $\Delta U(t)$:

$$\Delta U(t) = \int_0^\infty \omega \Delta I(\omega, t) d\omega. \qquad \text{Eq. (1)}$$

Here, $\omega = E - E_F$, $t$ is the delay time value, and $\Delta I(\omega, t)$ is the energy distribution curve of the difference image at $t$ [lower panels of Fig. 4(a)] integrated over the emission angle [-15º, +15º]. $\Delta U(t)$ is a good measure of the excess electronic energy deposited by the pump pulse (for example, see Ref. [38]), and in the particular case for BP, of the excess energy carried by the conduction electrons because the integral region of Eq. (1) is in the unoccupied side.

In Figure 5, we display $\Delta U(t)$ versus *t* in linear-linear (a), semi-log (b), and log-log plots (c). First, the recovery lasting for more than 100 ps is discerned in the linear-linear plot [Fig.5 (a)] as the finite intensity remaining even after 100 ps. In the semi-log plot [Fig. 5(b)], we observe that the decay cannot be fit by a single exponential function: At least three exponential terms are needed to have a reasonable fit to the decay profile. That is, the dissipation exhibits a non-exponential-type decay, or a slower-than-exponential-type decay. The non-exponential profile seen in Fig. 5(b) motivated us to display $\Delta U(t)$ in the log-log plot [Fig. 5(c)], which is convenient to judge whether a power-law-type decay, $\propto t^{-a}$, is occurring or not. The power of the decay after 100 ps is read to be $a \lesssim 0.05$, and is much smaller than any powers expected for the spatial dissipation of heat; namely, $a = 0.5$ and 1 for dissipations in one- and two-dimensions, respectively [38]. In fact, the time region ~100 ps is known to be still before the spatial diffusion of heat prevails. In this time region, $a \simeq 0.05 - 0.3$ is expected when the heat transfer between the high-energy phonons and low-energy phonons plays the key role in the recovery [39]. While the microscopic mechanism that dominates the long recovery profile still remains elusive [40], our analysis reveals that the longevity is unconventional and does not exclude the possibility that the underlying mechanism may incorporate the formation of the long-lived excitons.

**Conclusions**

In this study, we have observed a giant anisotropy in the effective masses of electrons (holes) in the conduction (valence) band along *c*- and *a*-axes of BP by ARPES and TARPES measurements. Moreover, we have also observed the photo-carriers generated across the direct band gap in BP maintained over 400 ps. This long duration of photoexcited electron at the bottom of the conduction band can be attributed to the stabilized e-h pairs between conduction and valence bands and also overcome graphene's short carrier lifetime induced constrains for the mid-infrared applications. The significant feature of excitonic insulating state is still lacking in the present study probably because of the insufficient temperature *i.e.* the predicted critical temperature is typically lower than 1K [10-11] and the larger probe energy *i.e.* the probe wavelength is better to match the band gap energy. But our experimental findings certainly provide precursor information for the excitonic condensates in optically pumped semiconductors and also paves the way for developing versatile broadband mid-infrared optoelectronic devices.

**Methods**

Single crystalline BP samples were grown by high pressure Bridgeman method as described elsewhere [41]. STM and ARPES measurements were conducted using an LT-STM (Omicron NanoTechnology GmbH) and the synchrotron radiation at the beamline (BL-7) equipped with a hemispherical photoelectron analyzer (VG-SCIENTA SES2002) of the Hiroshima Synchrotron Radiation Center (HSRC). The TARPES experiment was performed using the pump-and-probe configuration at the Laser and Synchrotron Research (LASOR) Center of Institute for Solid State Physics (ISSP), the University of Tokyo. The TARPES system equipped with an amplified Ti: sapphire laser system delivering 1.48 eV pulses of 170 fs duration with 250 kHz repetition and a hemispherical photoelectron analyzer (VG Scienta R-4000). The laser is split into two beams; one pulse was used as a pump while the other was up-converted into 5.92 eV and used as a probe. The energy resolution and the Fermi energy position ($E_F$) were determined by recording the Fermi cutoff of Au in electrical contact to the sample and the analyzer. The delay time between pump and probe pulses was tuned by a delay stage changing the optical path length of the pump beam line. The delay origin $t = 0$ and the time resolution of 280 fs were calibrated by using the pump-and-probe photoemission signal of graphite attached next to the sample [42]. Samples of BP were *in situ* cleaved by a Scotch tape along (010) plane for all measurements and measured at 5 K for ARPES, 77 K for STM in an ultra-high vacuum better than $1 \times 10^{-9}$ Pa. We confirmed that the *in situ* cleaved BP surface was free from oxidation and stable during the measurements by X-ray photoelectron spectroscopy (XPS) [see Figs. S2 and S3 of Supplementary Material].

**Acknowledgments**

The TARPES measurement was carried out by the joint research in ISSP, University of Tokyo. This work was partly supported by JSPS Kakenhi (Grant Nos. 26247064, 2680015, 17H06138, 18H01148 and 18H03683).

**Author contributions**

Y. A. grew the bulk black phosphorus single crystals. M.N., R.Y., Y.I., Z.S., and M.N., performed the experiments and analyzed the results. M.N. and A.K. wrote the manuscript with inputs from all authors. Y.U, M.T, S.S. and A.K. supervised work and discussed the results. All authors contributed to the scientific discussion and manuscript preparation.

**Competing financial interests**

The authors declare no competing interests.

**Figures**

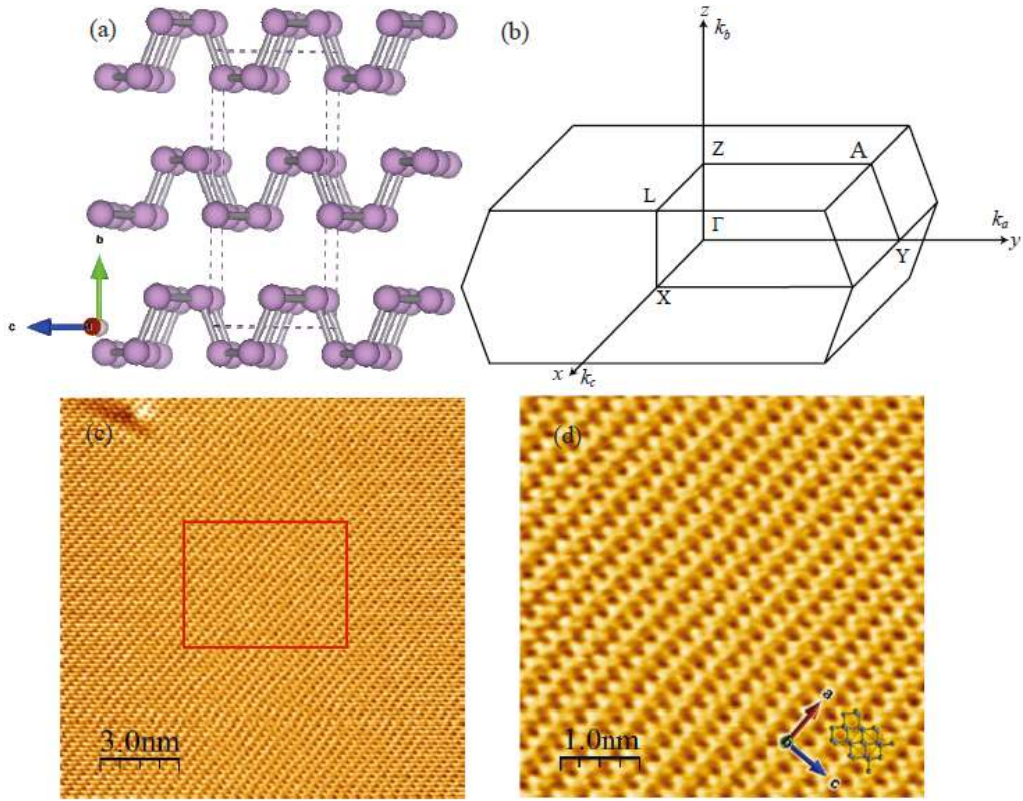

**Figure-1.** (a) Crystal structure of BP from side view. (b) Schematic Brillion Zone of BP. (c) Large-scale STM topography image of BP surface with scan size 15 nm×15 nm. (d) Closed-up image of the region corresponding of the selected area in c showing detailed features of the BP surface structure along *a*- and *c*- axes. Images were taken with sample-bias voltage of 1.0 V and a tunneling current of 0.15 nA at 77 K.

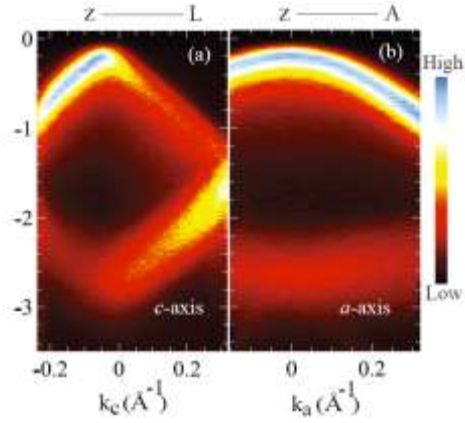

**Figure-2.** High-resolution ARPES spectrum along Z-L direction (a) and Z-A direction (b) with photon energy 19 eV.

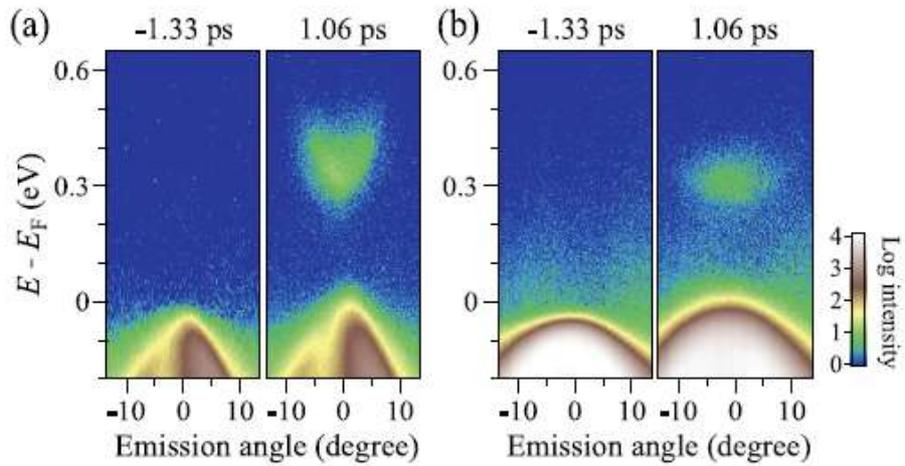

**Figure-3.** TARPES images of BP with valence and conduction bands along *c*-axis (a) and *a*-axis (b) recorded with before and after pumping. The pump-probe delay times and axis are noted on the spectra.

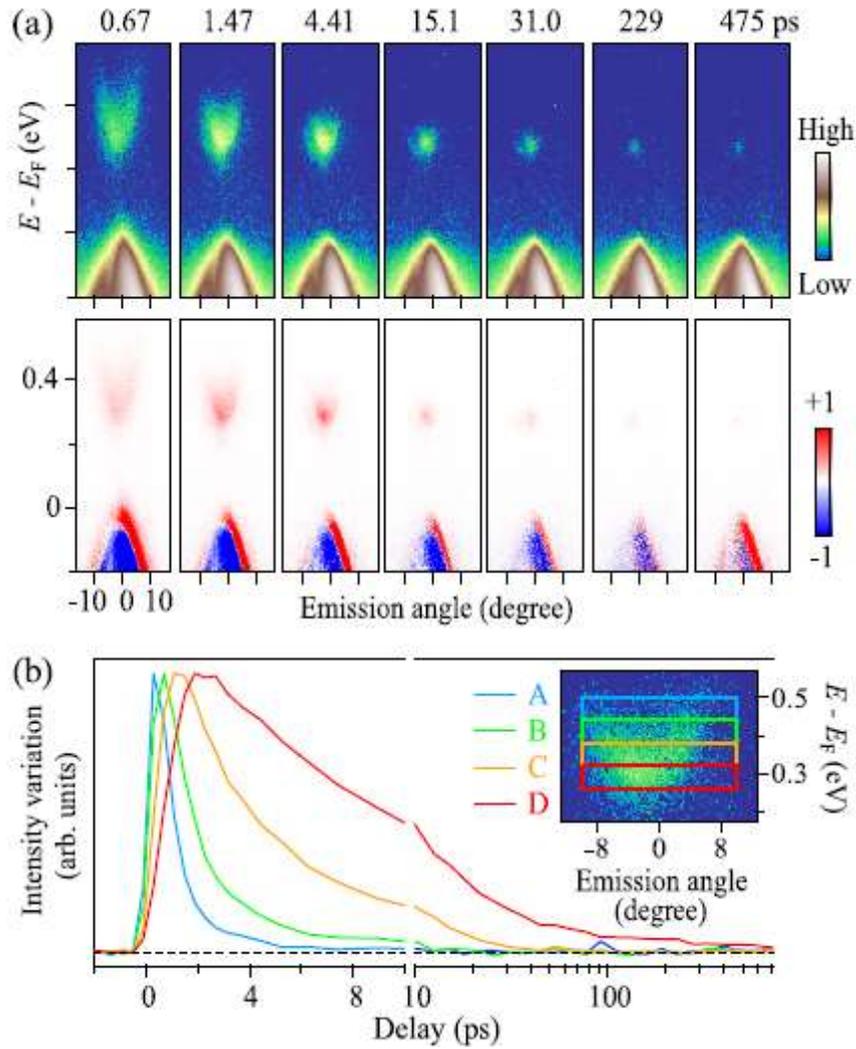

**Figure-4.** (a) TARPES spectra of BP along c-axis (Upper panels) recorded at a various pump-probe delay times and its difference to the image before pumped (Lower panels). (b) Ultrafast time evolution of conduction band population in frames A-D. The frames A to D represent integrated transient photoemission intensity in the angular range of (±8) degree and energy region of [0.44, 0.50], [0.38, 0.44], [0.32, 0.38], [0.26, 0.32] (units of eV), respectively.

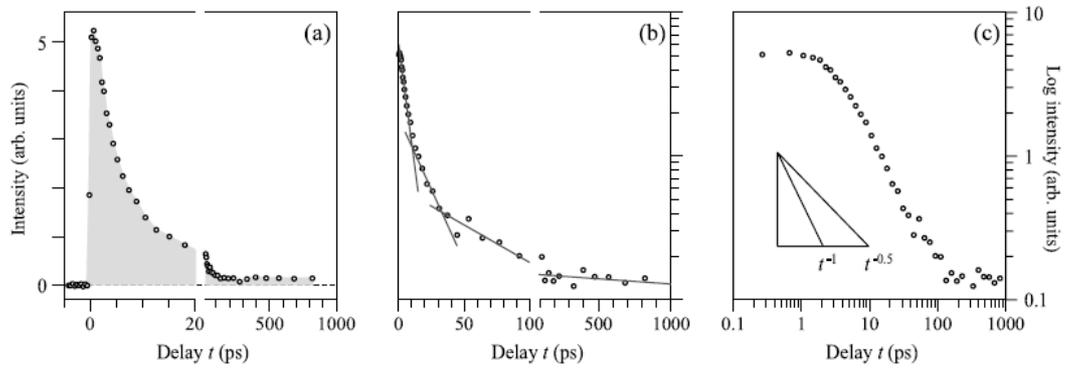

**Figure-5.** Recovery profile of the electronic energy. $\Delta U(t)$ versus $t$ plotted in linear-linear (a), semi-log (b), and log-log plots (c). The profiles for the power-law-type decay $\propto t^{-a}$ of $a = 0.5$ and 1.0 are displayed as slopes in (c).